\documentclass[preprint]{aastex}

\shorttitle{Comet Science with JWST}

\shortauthors{Kelley et al.}

\usepackage{amsmath}

\newcommand\efrho{$\epsilon f\rho$}
\newcommand\afrho{$Af\rho$}
\newcommand\athfrho{$A(\theta)f\rho$}
\newcommand\jameswebb{\textit{James Webb Space Telescope}}
\newcommand\jwst{\textit{JWST}}
\newcommand\hho{H$_2$O}
\newcommand\coo{CO$_2$}
\newcommand\chhhh{CH$_4$}
\newcommand\molecps{molecules~s$^{-1}$}
\newcommand\uJy{$\mu$Jy}
\newcommand\wmm{W~m$^{-2}$}
\newcommand\wmmm{W~m$^{-2}$~\micron$^{-1}$}

\newcommand\refeq[1]{Eq.~\ref{eq:#1}}
\newcommand\reffig[1]{Fig.~\ref{fig:#1}}
\newcommand\reftab[1]{Table~\ref{tab:#1}}
\newcommand\refsec[1]{Section~\ref{sec:#1}}

\begin{document}
\slugcomment{Accepted for publication in the Publications of the Astronomical Society of the Pacific, 2015 Oct 15}
\title{Cometary Science with the \jameswebb}

\author{
  Michael S. P. Kelley\altaffilmark{1},
  Charles E. Woodward\altaffilmark{2},
  Dennis Bodewits\altaffilmark{1},
  Tony L. Farnham\altaffilmark{1},
  Murthy S. Gudipati\altaffilmark{3,4},
  David E. Harker\altaffilmark{5},
  Dean C. Hines\altaffilmark{6},
  Matthew M. Knight\altaffilmark{7},
  Ludmilla Kolokolova\altaffilmark{1},
  Aigen Li\altaffilmark{8},
  Imke de Pater\altaffilmark{9},
  Silvia Protopapa\altaffilmark{1},
  Ray W. Russell\altaffilmark{10},
  Michael L. Sitko\altaffilmark{11},
  Diane H. Wooden\altaffilmark{12}
}

\altaffiltext{1}{Department of Astronomy, University of Maryland, College Park, MD 20742-2421, USA}

\altaffiltext{2}{Minnesota Institute for Astrophysics, 116 Church Street S. E., University of Minnesota, Minneapolis, MN 55455, USA}

\altaffiltext{3}{Science Division, Jet Propulsion Laboratory, California Institute of Technology, Mail Stop 183-301, 4800 Oak Grove Drive, Pasadena, CA 91109, USA}

\altaffiltext{4}{Institute for Physical Sciences and Technology, University of Maryland, College Park, MD 20742, USA}

\altaffiltext{5}{Center for Astrophysics and Space Sciences, University of California, San Diego, 9500 Gilman Dr., La Jolla, CA 92093-0424, USA}

\altaffiltext{6}{Space Telescope Science Institute, 3700 San Martin Dr., Baltimore, MD 21218, USA}

\altaffiltext{7}{Lowell Observatory, 1400 W. Mars Hill Rd, Flagstaff, AZ 86001, USA}

\altaffiltext{8}{Department of Physics and Astronomy, University of Missouri, Columbia, MO 65211, USA}

\altaffiltext{9}{Department of Astronomy, 501 Campbell Hall, University of California, Berkeley, CA, 94720, USA}

\altaffiltext{10}{The Aerospace Corporation, Los Angeles, CA 90009, USA}

\altaffiltext{11}{Space Science Institute, Boulder, CO 80301, USA}

\altaffiltext{12}{NASA Ames Research Center, Moffett Field, CA 94035-0001 USA}

\begin{abstract}

The \jameswebb{} (\jwst), as the largest space-based astronomical observatory with near- and mid-infrared instrumentation, will elucidate many mysterious aspects of comets.  We summarize four cometary science themes especially suited for this telescope and its instrumentation: the drivers of cometary activity, comet nucleus heterogeneity, water ice in comae and on surfaces, and activity in faint comets and main-belt asteroids.  With \jwst{}, we can expect the most distant detections of gas, especially \coo, in what we now consider to be only moderately bright comets.  For nearby comets, coma dust properties can be simultaneously studied with their driving gases, measured simultaneously with the same instrument or contemporaneously with another.  Studies of water ice and gas in the distant Solar System will help us test our understanding of cometary interiors and coma evolution.  The question of cometary activity in main-belt comets will be further explored with the possibility of a direct detection of coma gas.  We explore the technical approaches to these science cases and provide simple tools for estimating comet dust and gas brightness.  Finally, we consider the effects of the observatory's non-sidereal tracking limits, and provide a list of potential comet targets during the first 5 years of the mission.

\end{abstract}


\section{INTRODUCTION}
The large aperture of the \jameswebb{} (\jwst) in combination with its sensitive near-infrared (near-IR) and mid-infrared (mid-IR) instruments will provide new opportunities to study comet dust, gas, and nuclei at moderate and large heliocentric distances (defined here as beyond 3~AU), and at excellent spatial resolutions for closer objects.  Moreover, \jwst's operational strategies, scheduling, and instrument capabilities allow us to follow comets over a wide range of times and heliocentric distances.  Nominal performance expectations for \jwst{} hardware and flight software will enable the telescope to follow moving targets with apparent rates up to $\simeq 0\farcs030$~s$^{-1}$ with small pointing errors (currently specified to be 17~mas RMS at 0\farcs030~s$^{-1}$).  \jwst{} will use the JPL HORIZONS system ephemerides for pointing and tracking of comets with known orbital elements, although an observer may provide ephemerides for targets not in the JPL database. Accurate ephemerides and/or peak-up operations are essential to place comets and other Solar System small bodies in the narrow ($\lesssim1$\arcsec) spectrometer slits. Observing visits are limited to the dwell time of the guide star on the fine-guidance detectors, which have a field-of-view of $2.2 \times 2.2$ arcmin. Non-sidereal guiding strategies are beginning to be developed and refined by the \jwst{} team and will be verified during the initial observatory commissioning period. Release of moving target capabilities is anticipated to be at the start of routine science operations.  See \citet{norwood15} and \citet{milam} for summaries of Solar System science with \jwst{}.

The study of comets with \jwst{} will address current key decadal questions in planetary science, as well as astronomy and astrophysics, that are challenging to address with ground-based observations alone.  These questions pertain to the initial stages of solar system formation and its subsequent evolution, and include: ``What were the initial stages, conditions, and processes of Solar System formation and the nature of the interstellar matter that was incorporated?'' ``What were the primordial sources of organic matter, and where does organic synthesis continue today?'' ``How have the myriad chemical and physical processes that shaped the solar system operated, interacted, and evolved over time?'' \citep{ps-decadal13}.

Here, we highlight some of the observational challenges, experimental techniques, and representative science opportunities that \jwst{} offers as a tool to improve our knowledge of comets.  Such a space-based platform has easy access to key cometary emission bands from \hho, \coo, CO, and \chhhh{} that would otherwise be hindered by severe telluric absorption.  Except for \coo, studies of these latter molecules are possible from ground-based telescopes, e.g., through photodissociation products (e.g., OH and O from \hho, O from \coo), through non-resonance emissions (e.g., water hot-bands) that are not blocked by the Earth's atmosphere, or individual rovibrational lines Doppler-shifted from their terrestrial counterparts \citep[for a review, see][]{bockelee-morvan04}.  However, having direct access to the fundamental bands, and without Doppler shift restrictions, can enable new and exciting results.  Above the atmosphere, background emission is greatly reduced, allowing for greater sensitivity to the 3-\micron{} region and beyond, enabling survey studies of dust emission, water ice absorption, or (refractory) organic features that might otherwise take years to achieve from ground-based telescopes.  Time-domain \jwst{} observational campaigns covering a wide range of heliocentric distances become possible for many more targets, rather than just the occasional Hale-Bopp-class comet.

In \refsec{brightness}, we present our methods to estimate the continuum arising from dust (reflected and thermal emission), and to estimate the strengths of emission bands from molecules.  These methods are used to determine integration times and signal-to-noise ratios throughout our paper.  We then propose four science examples that \jwst{} will be especially suited to address.  In \refsec{gas-coma}, we discuss the telescope's ability to measure the primary drivers of comet activity: \hho, \coo, and CO.  In \refsec{dust}, we summarize simultaneous observations of water gas and dust comae and how they can be used to assess nucleus heterogeneity.  In \refsec{ice}, we propose observations that can address the nature of ice in cometary comae and nuclei.  Finally, in \refsec{faint}, we demonstrate how \jwst{} can be used to detect gas in main belt and other faint comets.  The effects of the observatory's non-sidereal tracking limit on observations of comets, and specific comet observing opportunities are summarized in \refsec{targets}.

\section{COMET BRIGHTNESS ESTIMATION TOOLS}\label{sec:brightness}
\subsection{Continuum}\label{sec:continuum}
At \jwst{} wavelengths ($0.6~\micron{}<\lambda<28.5$~\micron{}), comet spectra are generally dominated by sunlight scattered by and thermal emission from coma dust.  It is a challenge for observers to accurately predict the surface brightness of a given target, due to each comet's inherent variability and coma physical properties (dust and gas ejection speeds, grain parameters, presence of ice, etc.).  For comets in the inner Solar System, the observed energy from dust thermal emission and that from light scattered by dust are approximately equal in the 2--4~\micron{} range, thus a single Planck function or stellar template are poor approximations.  Whether the goal is to observe the continuum from dust, absorption or emissivity features, or gas emission, having a simple model to estimate continuum brightness is beneficial to observers interested in this wavelength regime.

Light scattered by comet dust can be estimated using the \afrho{} parameter of \citet{ahearn84-bowell}.  \afrho{} is the product of the albedo, $A$, the dust filling factor, $f$, and the radius of the circular aperture in consideration, $\rho$.  Under certain assumptions and conditions, it is proportional to the dust production rate \citep{fink12}.  It carries the units of $\rho$, and is typically expressed in cm,
\begin{equation}
  A(\theta)f\rho = \frac{4 \Delta^2 r_h^2}{\rho}\frac{F_{\lambda,c}}{F_{\lambda,\sun}}\mbox{ (cm),}
  \label{eq:afrho}
\end{equation}
where $\Delta$ is the observer-comet distance (cm), $r_h$ is the heliocentric distance of the comet (AU), $F_{\lambda,c}$ is the observed flux density of the comet within a circular aperture (\wmmm), $\rho$ is the radius of the aperture projected to the distance of the comet (cm), and $F_{\lambda,\sun}$ is the flux density of sunlight (\wmmm{AU$^{-2}$}).  We have written the albedo as $A(\theta)$ to emphasize that the observed albedo is a function of phase angle (Sun-comet-observer angle, $\theta$).  For observations corrected to a phase angle of 0\degr, we will simply write \afrho.  The albedo of \citet{ahearn84-bowell} is a factor of 4 larger than the geometric albedo, $A_p$, of \citet{hanner81-albedo}: $A(\theta)=4A_p(\theta)$.  Throughout the paper, we use the combined Halley-Marcus phase function from D.~Schleicher to scale \athfrho{} to \afrho{} \citep{schleicher98, marcus07b, schleicher11}.

For quantifying the thermal emission, \citet{kelley13-activity} introduced the parameter \efrho{}, which is the thermal emission equivalent to \afrho:
\begin{equation}
  \epsilon f\rho = \frac{\Delta^2}{\pi \rho}\frac{F_{\lambda,c}}{B_{\lambda}(T)}\mbox{ (cm),}
  \label{eq:efrho}
\end{equation}
where $\epsilon$ is the apparent emissivity, $F_{\lambda,c}$ is now the observed thermal emission from dust (\wmmm), and $B_{\lambda}(T)$ is the Planck function evaluated at temperature $T$ (\wmmm, and K, respectively).  Similar to \afrho{}, this quantity carries units of length, determined by the units of $\rho$.  \citet{kelley13-activity} recommend using the effective (color) temperature of the continuum, with a default temperature of $T=306\ r_h^{-1/2}~\mathrm{K}$, 10\% warmer than an isothermal blackbody sphere in local thermodynamic equilibrium with insolation ($T_{BB}=278\ r_h^{-1/2}~\mathrm{K}$).

For exposure time calculations, we propose using a combination of the two empirical quantities above:
\begin{equation}
  F_{\lambda,c} = A f \rho\ \frac{\Phi(\theta) \rho F_{\lambda,\sun}}{4 \Delta^2 r_h^2}
    + \epsilon f \rho\ \frac{\pi \rho B_{\lambda}(T)}{\Delta^2}, \mbox{ (\wmmm)}
    \label{eq:comet-model}
\end{equation}
where $\Phi(\theta)$ is the phase function of the dust evaluated at the phase angle $\theta$.  The filling factors for \afrho{} and \efrho{} are not necessarily the same, as different populations of dust could dominate the measured scattered and thermal emission.  In \reftab{ef2af} we present seven observations of five comets with contemporaneous \afrho{} and \efrho{} estimates.  The ratios $\epsilon{}f\rho / Af\rho=\epsilon f_{em} / A f_{sca}$ aspan the range 2.3 to 4.2.  When \efrho{} is not known, we suggest scaling \afrho{} by the ratio $\epsilon f_{em} / A f_{sca} = 3.5$, based on the observed range in our small sample, and the assumptions: (1) $\epsilon\approx0.85-0.95$, (2) $A\approx0.25$, and (3) the emission and scattering filling factors are similar, $f_{em}\approx f_{sca}$.  Because the \efrho{} parameter depends on effective temperature, the ratio $\epsilon{}f_{em}/Af_{sca}$ may also be correlated with $T/T_{BB}$, although this is not evident in the limited dataset in \reftab{ef2af}.  A caveat to our approach arises if $T/T_{BB}$ and \efrho{} rely on data measured over a limited wavelength range.  Any derived values may not be valid for other wavelengths, or even heliocentric distances.  Specifically, short ward of $\sim$7.5~\micron{} silicates have limited absorptivity so the 3--7~\micron{} continuum is dominated by a warmer continuum from more absorbing carbonaceous grains.

\begin{figure}
  \includegraphics[width=\textwidth]{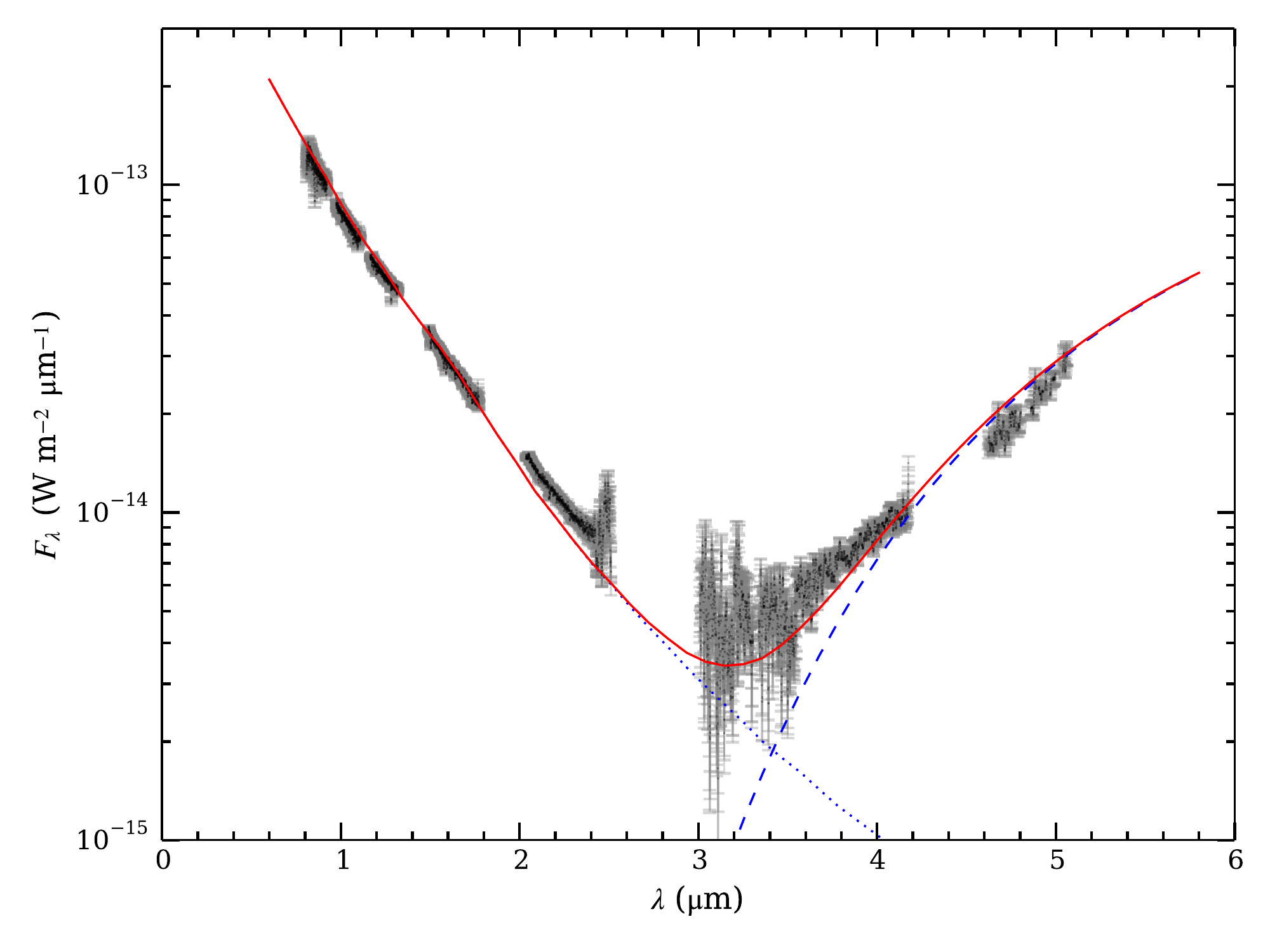}
  \caption{Spectrum of the coma of comet
    73P-C/Schwass\-mann-Wach\-mann~3 from \citet{sitko11} and a model
    fit using \refeq{comet-model} (solid line).  The dotted and dashed
    lines are the model scattered and thermal components,
    respectively. \label{fig:sw3}}
\end{figure}

\placetable{tab:ef2af}
\placefigure{fig:sw3}

In \reffig{sw3}, we plot a near-IR spectrum of comet 73P-C/Schwass\-mann-Wach\-mann~3 \citep{sitko11}, and our proposed model with the parameters $T/T_{BB}=1.12$, \afrho=1400~cm, and $\epsilon{}f/Af=3.0$.  The agreement is acceptable for first-order integration time estimations.  Further refinements, e.g., wavelength dependent \afrho{} to account for coma color, or wavelength dependent \efrho{} to account for grain composition and size distribution, are not necessary.

\subsection{Gas Flux}\label{sec:gas}
In addition to the dust continuum, gas emission bands are commonly observed.  Here, we summarize our method to estimate total band flux.  Using fluorescence $g$-factors (photons~s$^{-1}$~molecule$^{-1}$ at 1~AU) from \citet{crovisier83}, we generate total band fluxes in a 0\farcs4 radius aperture, for either the NIRSpec integral field unit or micro-shutter assembly, assuming an optically thin coma (\reftab{gas-flux}),
\begin{equation}
  F = \frac{Q \rho h c g}{8 \lambda v \Delta^2 r_h^2}\mbox{ (W~m$^{-2}$)},
  \label{eq:gas}
\end{equation}
where $Q$ is the production rate of the molecule (molecules~s$^{-1}$), $\rho$ is is the radius of the aperture projected to the distance of the comet, $h$ is the Planck constant, $c$ is the speed of light, $\lambda$ is the band central wavelength, $v=800\,r_h^{-1/2}$~m~s$^{-1}$ is the expansion speed of the gas, $\Delta$ is the observer-comet distance, and $r_h$ is the heliocentric distance of the comet (AU).  All parameters have MKS units, unless otherwise noted.

Although we assume an optically thin coma, optical depth effects must be carefully considered when interpreting real spectra.  For comet 9P/Tempel~1, \citet{feaga07} showed that optical depth effects are significant inside of $\sim$10~km for a water production rate of $5\times10^{27}$~\molecps{} and a CO$_2$ production rate of $5\times10^{26}$~\molecps{} at 1.5 AU from the Sun.  For integration time estimation purposes, we neglect line opacity effects, which should only affect total band fluxes at the 10\% level or less \citep[e.g.,][]{feaga14}. 

\subsection{Comet Comae and Exposure Time Calculators}
A \jwst{} exposure time calculator could provide a two-component model that would be adaptable to a variety of Solar System sources.  Continuum spectra of comet comae, comet nuclei, and asteroids can all be approximated by: (1) $F_{\lambda,refl}$, based on a solar-type spectrum, potentially reddened, representing the reflected or scattered light; and (2) $F_{\lambda,therm}$, a blackbody spectrum of arbitrary temperature, representing the thermal emission,
\begin{align}
  F_{\lambda,refl} &= C_{refl} \frac{F_{\lambda,\sun}(\lambda)}{F_{\lambda,\sun}(\lambda_{refl,0})}\\[2mm]
F_{\lambda,therm} &= C_{therm} \frac{B_\lambda(T,\lambda)}{B_\lambda(T,\lambda_{therm,0})}
\end{align}
where $C_{refl}(\lambda_{refl,0})$ and $C_{therm}(\lambda_{therm,0})$ are the reflected (or scattered) and thermal emission normalization factors; $F_{\lambda,\sun}$ is the solar flux density \citep[see, e.g.,][]{astm06}; and $B_\lambda$ is the Planck function.  The template spectra are normalized to 1.0 at reference wavelengths $\lambda_{refl,0}$ and $\lambda_{therm,0}$.  Because \afrho{} and \efrho{} can change with wavelength, their values and the normalization wavelengths should be carefully chosen for the observation in consideration.  For example, one might choose $\lambda_{refl,0}=1.0$~\micron{} and $\lambda_{therm,0}=4.0$~\micron{} for the spectrum in \reffig{sw3}.  The normalization factors for a comet coma model would be based the \afrho{} and \efrho{} components in \refeq{comet-model},
\begin{align}
  C_{refl} &= A f \rho \frac{\Phi(\theta) \rho}{4 \Delta^2 r_h^2} \\[2mm]
  C_{therm} &= \epsilon f \rho \frac{\pi \rho}{\Delta^2}\ .
\end{align}
Thus, comet-specific parameters would not need to be incorporated into an observing tool, but arbitrary scale factors would be a necessity.  Observers will need to estimate \afrho{} and \efrho{} based on the literature, their own experience, or recent observations.

\section{COMET SCIENCE}

\subsection{Comet Gas Coma Orbital Evolution}\label{sec:gas-coma}
Three gas species, \hho, \coo, and CO, are primarily responsible for driving cometary activity.  These molecules have very different levels of volatility \citep[e.g.,][]{langer00, meech04-activity, huebner06}, suggesting that each species should dominate the comet's activity at different parts of the orbit, delineated by the abundance of each ice, and the amount of solar energy available for sublimating those ices.  CO may be released at heliocentric distances out to tens of AU, while \coo{} sublimation dramatically increases around 6--8 AU and \hho{} sublimation near 2--3 AU.  However, the limited observations available for these species indicate that cometary behavior is more complex than would otherwise be suggested by a simple energy balance model.  The relative abundances not only differ from comet to comet, but also in unexpected ways with respect to heliocentric distance and sometimes before and after perihelion in a single comet \citep{bockelee-morvan04, mumma11-tax, ahearn12-origins, ootsubo12, feaga14, bodewits14-garradd}.  These variations are due presumably to inherent compositional differences between objects reflecting their origins in the early Solar System, heterogeneities in individual nuclei, and evolutionary processing from insolation.

Comprehensive data sets, i.e., those covering the production of all three molecules (\hho, \coo, and CO) over a wide range of times or heliocentric distances, typically depend on observations from multiple instruments and techniques.  How systematic uncertainties (instrumental, observational, or theoretical) affect the relative abundances is therefore of concern.  In addition, \coo{} cannot be observed from the ground, thus little is known regarding its abundance in comets or its role in comet activity outside of recent snapshot surveys \citep{crovisier00, ootsubo12, reach13} and missions to comets \citep{ahearn05, feaga07, ahearn11, hassig15}.  The wavelength coverage and sensitivities of NIRSpec, NIRCam, and NIRISS will allow \jwst{} to simultaneously or contemporaneously measure all three of the primary species over a range of heliocentric distances.  Such measurements enable studies of activity and composition at various distances and points around a comet's orbit, providing a uniform dataset that cannot be achieved by other means.  The high-spatial-resolution imaging available with NIRCam and NIRISS will also allow the investigation of heterogeneities in these species, revealed in the spatial distribution of the gases around the nucleus and their variations with time.

Here, we demonstrate the capabilities of time-domain spectroscopy with NIRSpec.  Our goal is to observe \hho, \coo, and CO in a single comet over a wide range of heliocentric distances.  We take the orbit of C/2013 A1 (Siding Spring) as a baseline (perihelion distance $q=1.4$~AU) to provide a rough idea for potential observing windows and comet brightnesses during an Oort cloud comet's journey from 10~AU pre-perihelion to 10~AU post-perihelion.  We assume a coma mixing ratio of 100/10/10 for \hho/\coo/CO at $r_h=1.5$~AU, approximately equal to the average observation listed in \citet{ahearn12-origins}.  Using the sublimation model of \citet{cowan79}, this ratio corresponds to equivalent active areas of 6~km$^2$ for \hho, 0.26~km$^2$ for \coo, and 0.080~km$^2$ for CO.  Other model quantities are summarized in \reftab{gas-model}.  For the sublimation model, we assume a rapidly rotating nucleus with a visual albedo of 0.05, an infrared emissivity of 1.0, and the pole direction perpendicular to the orbit.  Gas production rates are the product of the sublimation rate and the active area.

Water molecules have the shortest photodissociation length scale of the molecules in our study, but our apertures are no larger than 13\% of this scale, so for simplicity, we neglect photodissociation.  The continuum arising from dust, a source of noise, is modeled using \refeq{comet-model}, with the following parameters: $Af\rho=2000$~cm at 1~AU, scaling with $r_h^{-2}$, $\epsilon f_{em}/A f_{sca}=3.5$.

Production rates and \afrho{} values derived from the gas and dust models are listed at approximately 1~AU steps in \reftab{gas-flux}.  We have selected epochs that would be within \jwst's solar elongation constraints (85\degr{} to 135\degr).  The water production rate  ($9\times10^{27}$~\molecps) and \afrho{} (925~cm) at 1.5~AU are within a factor of two of actual measurements of the comet near perihelion \citep{schleicher14-cbet4004, bodewits15}.  Comet Siding Spring's \afrho{} at 4~AU (pre-perihelion) observed by \citet{li14-css} is higher than our model by an order of magnitude (120~cm versus 2000~cm).  Dynamically new comets tend to have shallow lightcurves upon approach to perihelion \citep{oort51, whipple78}, and we do not incorporate this aspect into our model.

\begin{figure}
  \includegraphics[width=\textwidth]{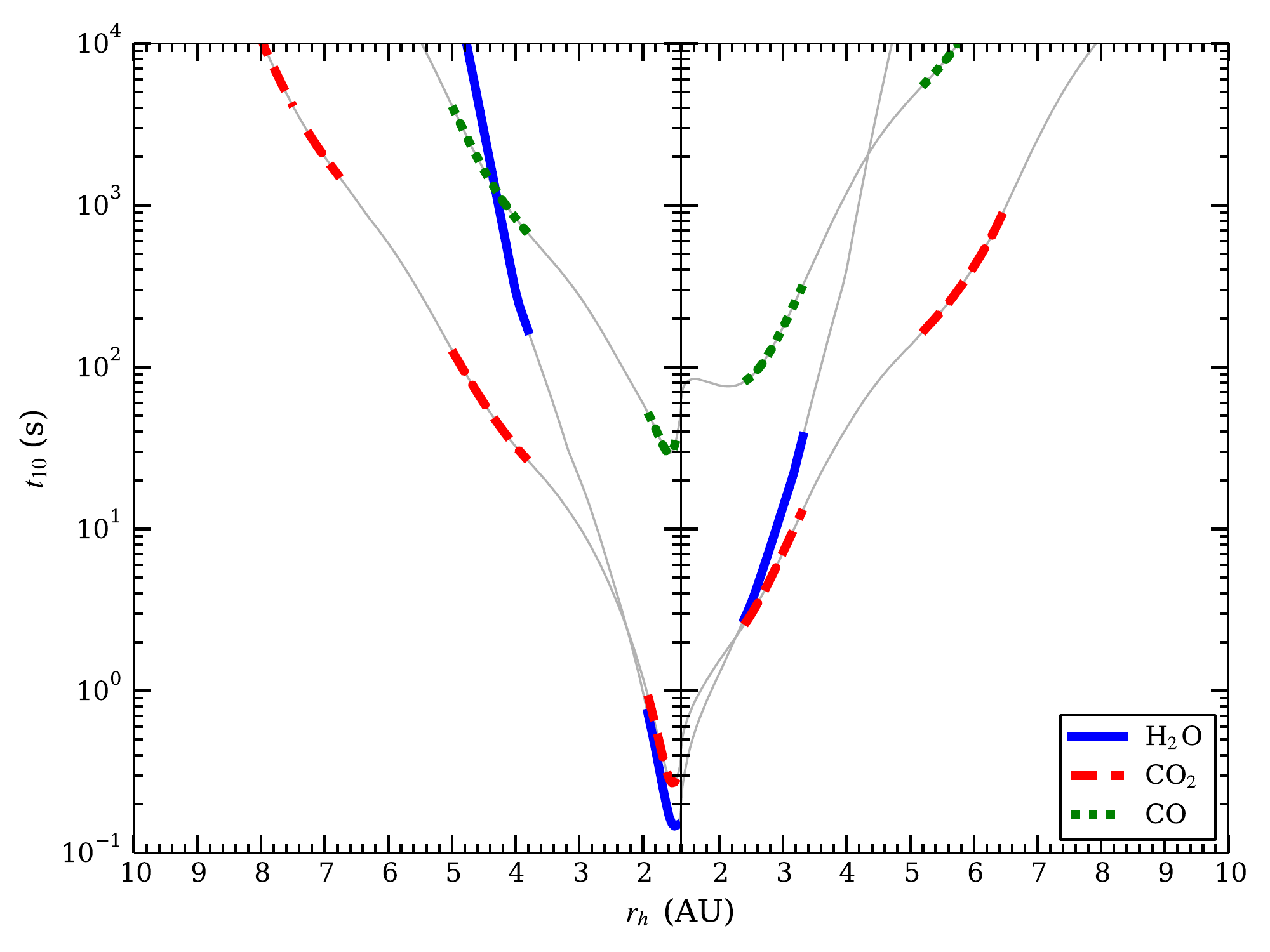}
  \caption{Estimated NIRSpec integration times (low-resolution mode) to achieve a signal-to-noise ratio of 10 at the band peak for each of \hho, \coo, and CO, given our example model described by \refsec{gas-coma} and \reftab{gas-flux}.  Thick lines mark the fictitious \jwst{} observing windows (solar elongation between 85\degr{} and 135\degr).   On the left are pre-perihelion epochs, on the right are post-perihelion epochs.\label{fig:gas}}
\end{figure}

Using the prototype NIRSpec exposure time calculator, we compute the integration time required to achieve a peak signal-to-noise ratio (SNR) of 10 for each band (\reftab{gas-flux}) and plot it in \reffig{gas}.  For our low-resolution spectroscopy observations, we assumed each band may be approximated by a single line with a full-width at half-maximum of 0.2~\micron.  Both \hho{} and CO are detected in a few thousand seconds at production rates $Q\gtrsim5\times10^{25}$~\molecps{} out to $\sim$4.5~AU; \coo{} is detected in a few thousand seconds at $Q\gtrsim1\times10^{25}$~\molecps{} out to $\sim$7~AU.  Thus, \jwst{} will easily be the best telescope for spectroscopic studies of comet activity at moderate and large heliocentric distance.

\subsection{Comet Dust Heterogeneity With MIRI}\label{sec:dust}
Comet nuclei are aggregates of the planetesimals present in the outer Solar System during the epoch of planet formation \citep{weissman04, belton07}.  There is an open question concerning the homogeneity of accreted planetesimals for particular comets, i.e., are individual comets aggregates of a heterogeneous population of planetesimals formed over a range of times and distances \citep[e.g.,][]{weidenschilling97-comets}?  Are comets collisional fragments \citep[e.g.,][]{stern01, morbidelli15} and less likely to be heterogeneous despite their origins?  High spatial resolution images and spectral data cubes from flyby spacecraft have suggested comet volatiles are not uniformly distributed about nuclei \citep[e.g.,][]{feaga07, ahearn11}, but the evidence for regions heterogeneous in dust properties relies on poorer resolution telescopic data, usually through multiple observations of a rotating comet \citep[e.g.,][]{ryan91, wooden04}.  A close approach (here, $\Delta<1$~AU) between a comet and \jwst{} would allow us to observe coma features on $\lesssim200$ km spatial scales, potentially revealing distinct active regions before they spatially mix in the coma.  Such a study can test the heterogeneity of comet nuclei by observing and comparing multiple spatially resolved coma jet features.

\jwst's MIRI instrument will be able to simultaneously observe both coma dust and water gas.  At  $\lambda > 5$~\micron, mid-infrared spectroscopy reflects a coma's dust composition, grain sizes, and porosities through solid-state emission features and thermal emission pseudo-continuum.  We use the terms grain and aggregate interchangeably.  They represent an entire particle of dust, which may be composed of smaller sub-units (the latter is implied for aggregate).  Typical dust features include a broad plateau from 8 to 12~\micron{}, primarily arising from anhydrous amorphous silicates (non-stochiometric olivine and pyroxene), and narrow emission peaks ($\lambda/\Delta\lambda \sim 10-20$) from crystalline silicates, dominated by Mg-rich olivine \citep{wooden02, hanner04}.  MIRI also covers the $\nu_2$ water gas emission band at 6~\micron{}, observed and studied in several comets with the \textit{Infrared Space Observatory} and \textit{Spitzer Space Telescope} \citep{crovisier97-iso-workshop, woodward07, bockelee-morvan09}.  In addition, the NIRSpec instrument holds the potential to simultaneously detect emission from water gas and hydrocarbons, the latter via the aromatic C--H stretch at $\sim$3.3~\micron{} and the aliphatic C--H stretch at $\sim$3.4~\micron{} \citep[see][]{li09}.

We consider comet 46P/Wirtanen as an example exercise demonstrating the technical feasibility of an observing program designed to test coma heterogeneity.  This comet has a close approach to \jwst{} in December 2018.  In November 2018 the comet is within the solar elongation constraints (85\degr{} and 135\degr) and quite bright, with integrated flux densities of order 1~Jy at 13~\micron{} (0\farcs5 radius aperture).  But rather than consider an observing window so soon after \jwst's launch (October 2018), we instead consider the post-closest approach window starting 2019 March 13 ($r_h=1.56$~AU, $\Delta=0.69$~AU) with non-sidereal rates of $\lesssim53$\arcsec~hr$^{-1}$, well under the \jwst{} limit of 108\arcsec~hr$^{-1}$.

Spatial-spectral maps of a comet's dust coma can be performed with either the Low Resolution Spectrometer (LRS, $\lambda/\Delta\lambda\sim100$) or Medium Resolution Spectrometer (MRS, $\lambda/\Delta\lambda\sim3000$) modes of MIRI.  Typically, comet dust studies would use the LRS, because the spectral resolving power of the MRS is much greater than necessary to resolve solid-state emission features.  However, we will design an observation for the MRS integral field units (IFU), which can efficiently map a $3\arcsec\times4\arcsec$ field of view, and have a spectral resolving power appropriate for measuring individual lines in the $\nu_2$ water band.

For our model, we adopt an effective nucleus radius of 0.6~km \citep{lamy98}, a maximum \afrho{} of 380~cm \citep{farnham98}, and a maximum water production rate $Q($\hho$)=1.6\times10^{28}$~\molecps{} at perihelion $q=1.05$~AU \citep{bertaux99}.  We also assume that the dust production rate scales with water production as $r_h^{-4.9}$ \citep{bertaux99}.

The IFUs can obtain complete spectra from 5 to 28.5~\micron{} in 3 exposures over a $3\farcs0\times3\farcs9$ field of view, corresponding to 1500~km~$\times$ 1900~km at the comet, with a spatial resolution of about 100~km per resolution element.  For comparison, the current best spatial resolution of this comet obtained with mid-infrared spectroscopy is 1300~km with \textit{Spitzer}/IRS.

Comet Wirtanen's peak surface brightnesses estimated for the MRS IFUs are listed in \reftab{miri-est}.  The surface brightness is an average within a 0\farcs4 radius aperture (corresponding to approximately 10 resolution elements at 6~\micron{} and a few resolution elements at 23~\micron{}).  Using the estimated MRS line sensitivities (in units of W~m$^{-2}$~pixel$^{-1}$) and dividing by the unresolved line width for each module yields continuum sensitivity thresholds  in units of flux density.  These values are scaled by $t^{-1/2}$ to an integration time of 100~s and listed in \reftab{miri-est}.  We expect high signal-to-noise ratios in the central aperture, easily enabling mapping of the inner coma.

For the \hho{} band at 6~\micron{} we assume an effective $g$-factor of $2.4\times10^{-5}$~photons~s$^{-1}$~molecule$^{-1}$ at 1 AU for the brightest lines, based on modeled \textit{Spitzer} spectra of C/2003~K4 (LINEAR), 71P/Clark, and C/2004~B1 (LINEAR) \citep{woodward07, bockelee-morvan09}.  These lines will be affected by optical depth effects, but introducing opacity is not necessary for our first-order feasibility exercise.  Following \refeq{gas}, we predict a total line flux $\sim2\times10^{-18}$~W~m$^{-2}$ in a 0\farcs4 radius aperture, or about $2\times10^{-19}$~W~m$^{-2}$~pixel$^{-1}$.  The lines will be easily detected above the continuum ($\sim10^{-20}$~W~m$^{-2}$~pixel$^{-1}$) with the MRS IFUs.  The morphology of the water gas would be used to locate active areas on the nucleus, and to identify dust that may originate from those areas.  Additional context with \coo{} and CO may be obtained with immediate follow-up observations with the near-IR spectrometer.  Comparisons of a particular active area's dust composition to that of the ambient coma or of other regions can be used to test the heterogeneity of the nucleus.

\subsection{Water Ice in Comae and on Nuclei}\label{sec:ice}
As preserved leftovers from the formation of the Solar System, comets serve as one of the best probes for studying primitive water ice in the form it may have existed in the early outer Solar System.  Thus observations of water ice in comets may allow us to test the accretion processes that led to the formation of comet nuclei and therefore of planets \citep[e.g.,][]{greenberg99}.  Contrasting coma ice properties to observations of ice on comet surfaces will allow us to study coma-nucleus interactions, and nucleus surface evolution.

Water ice has absorption bands in the near-infrared at 1.5, 2.0, and 3.0~\micron. The relative strength and shape of these features provides information on aggregate porosities, presence of dust within the ice, total abundance, and the size of the scattering units (following \refsec{dust}, we use the term grain and aggregate interchangeably).  Laboratory measurements show that infrared water-ice absorption bands change position and shape as a function of phase (crystalline or amorphous) and temperature: e.g., crystalline water ice  differs from its amorphous counterpart by the presence of a sharp and narrow feature at 1.65~\micron{} apparent at $T\lesssim200$~K \citep{grundy98}. The best approach to investigate water-ice physical properties is through near-infrared (1--5~\micron) spectroscopy.

Water ice has been observed in comae using ground- and space-based telescopes and flyby spacecraft \citep[e.g.,][]{davies97, kawakita04-ice, yang09-holmes, protopapa14}.  The small sample of comae detections presently available has revealed a range of water-ice characteristics: \micron{} and sub-\micron{} water-ice particles have been identified, as well as one case of crystalline water ice \citep[e.g.,][]{yang09-holmes, yang10-iauc, yang14, protopapa14}. However, the limited number of detections prevents development of a classification scheme or taxonomy based on water-ice properties.  The sample population is too small to apply statistical analyses with confidence, thereby limiting our understanding of the origins of this diversity.  It may relate to the different mechanisms responsible for delivering water ice from the interior into the coma (sublimation of more volatile ices e.g., \coo, CO, or through large outbursts in activity).  Alternatively, it may be the outcome of a preserved comet-to-comet heterogeneity.  Establishing these connections is crucial if water-ice properties are to be used as observationally imposed boundary conditions on comet nucleus formation models.

On nucleus surfaces, however, water ice has only been observed with flyby and orbiting spacecraft.  The observations of ice on the surfaces of comets 9P/Tempel~1 and 103P/Hartley~2 point towards sub-units with sizes of order 10 to 100~\micron{} \citep{sunshine06, sunshine12}.  This is much larger than the coma ice summarized above, and the ice excavated from the interior of Tempel~1 \citep{sunshine07}.  Thus, \citet{sunshine06} hypothesize that this ice is more likely due to re-condensation of water gas, rather than recently uncovered interior ice.  Observations of bare nuclei can help test when this re-condensation occurs: does it require constant replenishment from an active source, or is it deposited by low activity levels as a comet recedes from the Sun?  Complicating matters, \citet{capaccioni15} report the detection of a broad absorption band at 2.9 to 3.6~\micron{} in \textit{Rosetta} observations of the surface of comet 67P/Churyumov-Gerasimenko.  The absorption band has a peak depth of 20\% located at 3.2 to 3.3~\micron.  They suggest this feature is due to the presence of aromatic and aliphatic C-H bonds, carboxylic groups, and/or alcoholic groups.  The absorption feature has not been reported on the surfaces of 9P/Tempel~1 or 103P/Hartley~2 \citep{sunshine06, ahearn11}.  Whether absorption features are due to water ice or organic molecules, the study of comet surfaces at 3--4~\micron{} is best addressed through space-based missions or observatories.

NIRSpec is ideal for the study of water ice in comets.  The prism mode covers 0.7 to 5~\micron{} at a spectral resolving power of $R\sim100$, ideal not only to detect the broad absorption features of surface and coma ice but also to determine its chemical phase (amorphous or crystalline).  The instrument's higher spectral resolving powers ($R\sim1000$) may be sensitive to trapped \coo{} or other volatiles.  Below we summarize observations of a distant comet nucleus and coma with this instrument to illustrate future science opportunities.

\subsubsection{Surface ice}
At 5~AU and beyond, water ice deposited on the cold surface of a comet could persist until vigorous sublimation returns when surface temperatures rise to 160~K or more.  Where that ice resides would depend strongly on deposition and insolation history, including latitude and surface roughness considerations.  Cooler nuclei in the outer Solar System, e.g., cometary Centaurs, may have surface ice present throughout their orbit.  As an example of NIRSpec's capabilities with respect to spectroscopy of distant nuclei, we consider comet 172P/Yeung.  This comet was observed to be a point source by \textit{Spitzer} \citep{kelley13-activity} at $r_h=4.25$~AU (pre-aphelion), with an effective nucleus radius of 5.7~km \citep{fernandez13}.  We model the comet nucleus thermal emission with the near-Earth asteroid thermal model of \citet{harris98} and the parameters of \citet{fernandez13}.  The scattered light contribution is modeled assuming a geometric albedo of $A_p=0.04$ \citep{lamy04}.  Our goal is to detect the presence of water ice on a surface.  We choose the 3-\micron{} water ice because of its high relative absorption compared to the other water ice bands in the near-infrared.  The same methodology could be applied to the detection of surface organics, similar to those seen at comet 67P/Churyumov-Gerasimenko \citep{capaccioni15}.  There is, however, an important caveat: observations of a point source are not necessarily those of a bare nucleus, as a faint and/or unresolved coma may be present.  Supporting observations may be needed to rule out coma contamination in any particular data set.

We target a water-ice band with a band depth of 10\% or more at a confidence level of 5-$\sigma$.  Averaging over a 0.3-\micron-wide band (2.9 to 3.2~\micron), and taking a spectral resolving power of 100, we set our signal-to-noise ratio goal to 20 per resolution element.  \jwst{} can observe comet Yeung on 2019 June 30 ($r_h=4.5$~AU, $\Delta=4.4$~AU, phase angle $=13$\degr), when the comet's 3-\micron{} flux density would be approximately 5.7~\uJy{} (the thermal contribution is $<$5\% at $\lambda<$4~\micron).  We meet our SNR goal with NIRSpec's low-resolution, fixed-slit mode in 300 seconds of integration time.  Since the integration time is short compared to all known comet rotation periods ($>$5 hr), longitudinally resolved observations would be possible, providing important motivation to measure this comet's rotation period by 2019.

The Centaur 95P/Chiron (2060) is known to have water ice and activity \citep{hartmann90, luu90, foster99, luu00, duffard02, ruprecht15}, and a spectrum taken in Jul 2019 ($r_h=19$~AU, $\Delta=18$~AU, $F_\nu=21~\mu$Jy for $A_p=0.15$ and 80~km radius, \citealt{campins00}) could easily observe a 3-\micron{} feature as weak as a few percent.  In 1.0~hr, an $R\sim1000$ spectrum from 2.9 to 5.0~\micron{} can be obtained with a 3~\micron{} SNR of 60 per resolution element.  This higher resolution spectrum, and complementary spectra at shorter wavelengths, could be used to investigate the structure and composition of the ice (crystallinity, dirt fraction, trapped volatiles), as well as to look for signatures of organics in the 3- to 4-\micron{} region.

In an hour of integration time, we can achieve a SNR of $\sim$20 on Centaurs as small as $\sim$12~km radius at the distance of Chiron, sufficient for detecting a 3-\micron{} absorption feature at the 10\% level.  Observations of a representative sample of Centaur surfaces within \jwst's lifetime will be possible.  These observations of Centaurs, as transitional objects between the trans-Neptunian and Jupiter-family comet regions, will yield new insight into the evolutionary processes currently at work in our Solar System.

\subsubsection{Water Ice in Comet Comae}
For our comet coma case, we move beyond ice detection and instead focus on physical characterization.  A signal-to-noise ratio of $>50$ is necessary to ensure that our observations are sensitive to water-ice fractions of a few percent by area, to constrain the size of the scattering units, and to test the presence of the potentially weak crystalline water-ice feature at 1.65~\micron.

As an example case we consider the observations of comet C/2012 S1 (ISON) by \citet{li13-ison} with the \textit{Hubble Space Telescope} when the comet was at 4.15 AU from the Sun and 4.24 AU from Earth. The authors report an \athfrho{} of 1300 cm at 0.6~\micron{} within a distance of 5000 km from the nucleus, which corresponds to a flux density of 50 and $\sim20$~\uJy{} at 0.6 and 3~\micron, respectively, within an angular diameter of 0\farcs4. We conclude that in 900~s we can obtain a spectrum of comet ISON with a SNR of 100 using the current NIRSpec prototype exposure time calculator.  \citet{li13-ison} report reddening of the dust coma between 5000 km and 10000 km from the nucleus, compatible with the presence of icy grains close to the nucleus. It would be interesting to test this hypothesis by measuring the coma ice distribution with NIRSpec.  The flux density at 10000~km from the nucleus at 3~\micron{} is estimated to be 0.4~\uJy. An exposure time of 3~h yields a SNR of 30, high enough quality to test for the presence and size of water ice grains, and to study its evolution from the inner-coma out to 10000 km.  This exercise demonstrates that JWST will be able to measure ice properties throughout the coma of moderately bright comets at $r_h\lesssim4$~AU.

\subsection{Activity in Faint Comets and Main-Belt Asteroids}\label{sec:faint}
\jwst's capabilities will transform our understanding of faint and/or weakly active comets by enabling observations that were heretofore at the limit of experimental techniques.  Such comets comprise several categories:
\begin{enumerate}

\item Main-belt comets (MBCs).  Activity recurring from orbit to orbit has been identified in a handful of MBCs \citep[e.g.,][]{hsieh06, hsieh11-238p, hsieh15}, but the mechanism for this activity is unknown.  To date, there have been no successful searches for signatures of volatiles on MBCs.

\item Nearly inactive and dormant comets.  A large number of suspected dead comets (inactive objects in comet-like orbits) have been identified in the asteroid population \citep[e.g.,][]{jewitt05, jenniskens08}, and occasionally asteroids show cometary activity.  For example, \textit{Spitzer}/IRAC observations of ``asteroid'' (3552) Don Quixote exhibited signs of \coo{} outgassing \citep{mommert14}.  In addition, there are known comets with extremely weak or only sporadic activity, e.g., 209P/LINEAR~41 \citep{schleicher14-cbet3881} or 107P/Wilson-Harrington \citep{fernandez07}.  High-sensitivity observations of volatiles can help confirm or better characterize their cometary natures.

\item Very small comets.  The population of small near-Sun comets are generally thought to be $<50$~m in radius \citep{knight10}, too small or weakly active to be observed beyond the solar coronagraphs on SOHO or STEREO.  Little is known about these small bodies, including what drives their activity: normal cometary activity via sublimation of ices or the sublimation of dust and refractory grains due to the extreme temperatures experienced near perihelion \citep[equilibrium temperatures exceeding 1000~K;][]{mann04}.

\item Distant comets.  Many comets show continuous activity outside the so-called water-ice line.  Their activities may be driven by the more volatile ices \coo{} or CO \citep{meech04-activity, meech13-ison, stevenson15}, or by the exothermic annealing of water ice from amorphous to crystalline states \citet{meech09, jewitt09-centaurs}.  Owing to their large distances, studies of activity are generally limited to observations of dust comae and tails.

\end{enumerate}
A common theme in the above cases is faint activity.  NIRCam imaging would be capable of searching for signs of very weak activity (coma and/or tail, both dust and gas) and of efficiently characterizing nucleus properties such as size, elongation, albedo, and beaming parameter.  NIRSpec might be able to spectroscopically detect the driving volatiles at production rates lower than previously achievable.  The comet-asteroid transition is currently poorly understood, and such results, whether of severely processed sun-grazing/sun-skirting comets, dormant comets, distant comets, or of activated asteroids, would yield insight into the ongoing evolutionary processes of our solar system.

To demonstrate \jwst's capabilities for characterizing activity in faint comets, we consider observations of main belt comet P/2010 R2 (La Sagra) in late Aug 2021 ($r_h=2.7$~AU, $\Delta=1.9$~AU).  This observing window is post-perihelion, during a period when comet La Sagra has been observed to be active with an \afrho{} near 50~cm \citep{hsieh12-lasagra}.  Estimates of the dust production rate by \citet{moreno11-lasagra} and \citet{hsieh12-lasagra} span from 0.1 to 4~kg~s$^{-1}$.  Assuming a dust to gas mass production ratio of 1:1, and assuming water sublimation is driving the activity, this corresponds to a water production rate with an order of magnitude of $10^{25}$~\molecps.

For this exercise, we use NIRSpec to detect the $\nu_3$ water band at 2.7~\micron, one of the strongest water bands in comets \citep{crovisier83} and one that cannot be observed from the ground \citep{bockelee-morvan04}.  In a small aperture of radius 0\farcs2, the estimated continuum flux density at 2.7~\micron{} is $1.7\times10^{-18}$~\wmmm{}, and the water band flux is $2.3\times10^{-20}$~\wmm{}.  With the NIRSpec prototype exposure time calculator and the $R\sim100$ prism, we model the water band as a single 0.2-\micron{}-wide line.  In a 1~hr exposure, the peak of the line is detected above the continuum with a signal-to-noise ratio of 4.  Thus, we can expect \jwst{} to provide one of the best tests for the drivers of main-belt comet activity.  

\section{NON-SIDEREAL RATES AND POTENTIAL TARGETS}\label{sec:targets}

\begin{figure}
  \includegraphics[width=\textwidth]{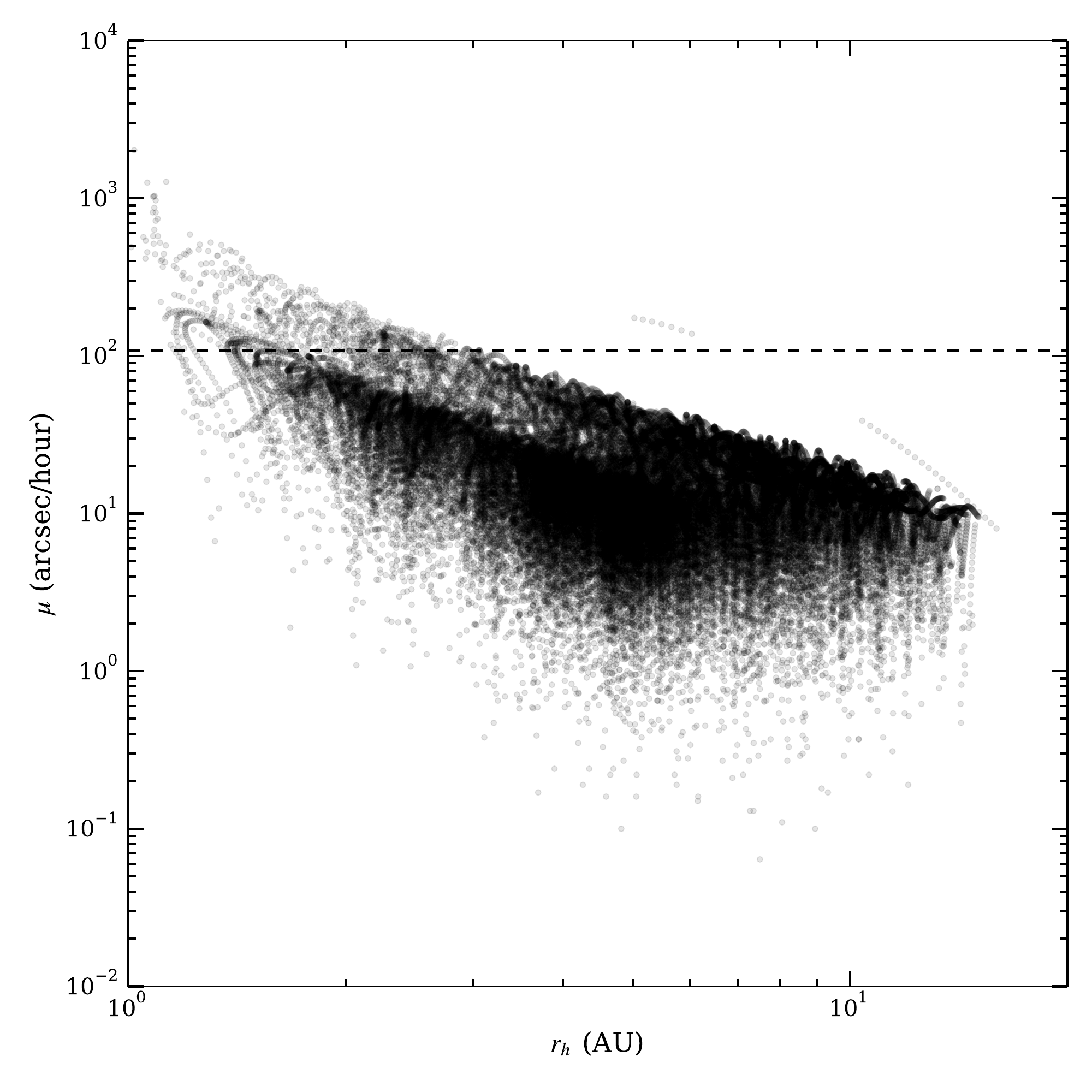}
  \caption{Comet proper motion ($\mu$) versus heliocentric distance ($r_h$) for all comets in our non-sidereal rate study.  Only epochs within \jwst's solar elongation constraints (85--135\degr) are shown.  A dashed horizontal line marks the observatory's current non-sidereal rate limit of 108\arcsec~hr$^{-1}$.  The isolated comet with high proper motion at 5 AU is P/2011~P1 (McNaught).  This object was potentially in a large orbit before passing Jupiter from a distance of 0.0008 AU in December 2010, and is currently in a Jupiter-family comet orbit. \label{fig:proper}}
\end{figure}

\jwst{} is expected to have the ability to track moving targets with proper motions $\leq$108\arcsec~hr$^{-1}$ (30 mas~s$^{-1}$) \citep{norwood15}.  To estimate the impact this limit may have on cometary science, we searched the NASA Jet Propulsion Laboratory Solar System Dynamics Group database for all comets that reached perihelion during the five-year period 2010 Jan 1 to 2015 Jan 1.  This list includes 393 comets: 221 short-period comets (with numbered or provisional ``P/'' designations), and 172 Oort cloud and long-period comets (with provisional ``C/'' designations).  For each target, we used the IAU Minor Planet Center's ephemeris service to generate ephemerides with 3-day intervals, using the center of the Earth as the observer.  The differences in ephemerides between the Earth and \jwst{} at the L2 Lagrange point (1.5$\times$10$^6$~km or 0.01~AU away) are not significant for this exercise.  The proper motions of all targets within \jwst's solar elongation constraints (85--135\degr) are shown in \reffig{proper}.  Not all of the observing windows make sense in a real world situation, e.g., the plot includes epochs before some comets are discovered, and all epochs are shown independent of the comet's apparent brightness.  However, we neglect these and other considerations and focus on the orbital characteristics of the known comet population.

\begin{figure}
  \includegraphics[width=0.5\textwidth]{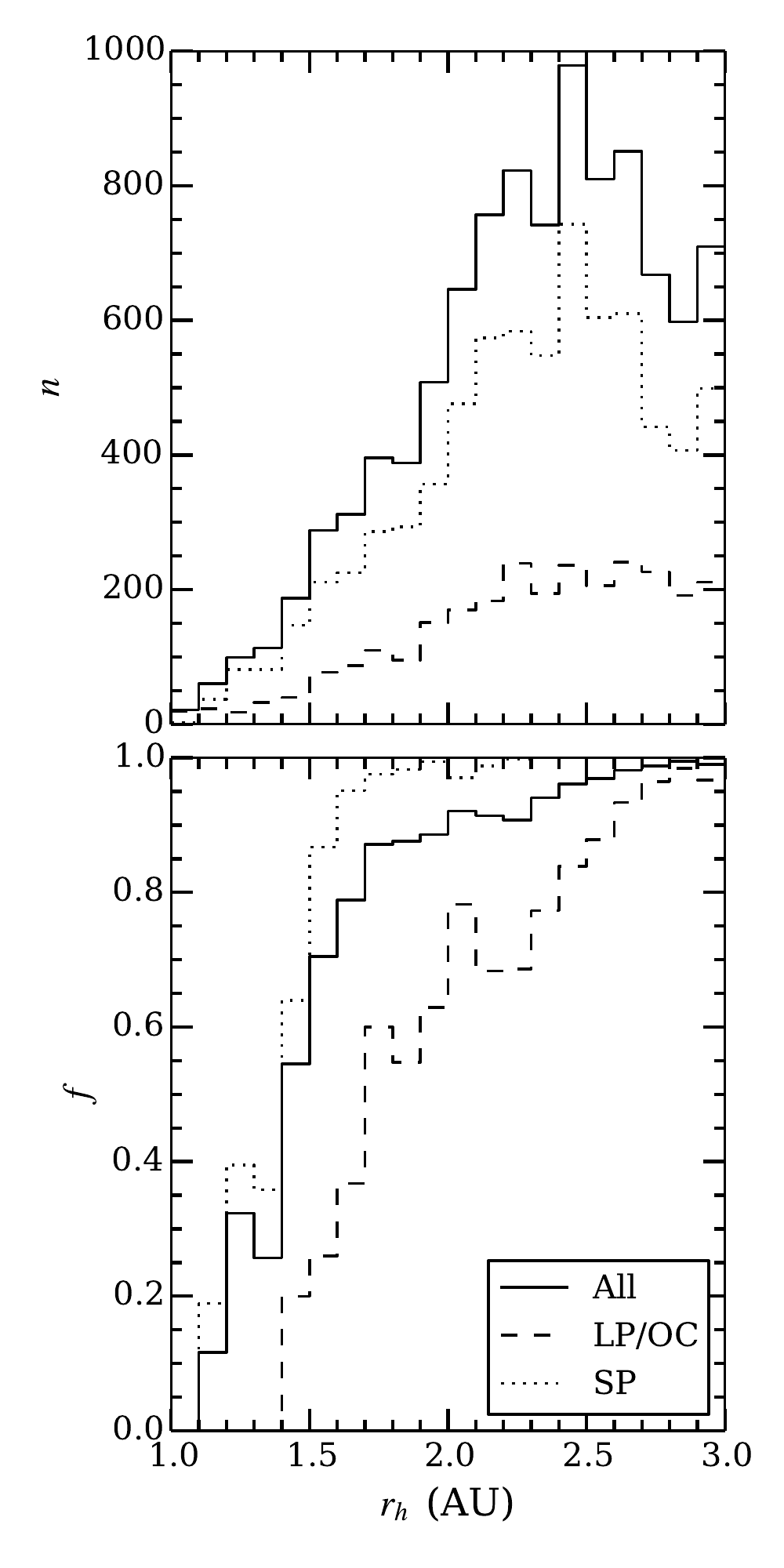}
  \caption{Top: A histogram of the number of epochs within \jwst's elongation constraints, $n$, versus heliocentric distance, $r_h$, for our non-sidereal rate study.  The solid line includes all comets, the dotted line only considers short-period comets, and the dashed line is limited to long-period and Oort cloud comets.  Bottom: The fraction of epochs with non-sidereal rates within \jwst's observing limits, $f$, versus heliocentric distance.   In general, short period comets are more accessible to \jwst{}, and the observatory's tracking limit significantly impacts observations within $r_h\lesssim 2.0$~AU. \label{fig:proper-hist}}
\end{figure}

In Fig.~\ref{fig:proper-hist}, we show histograms of our comet ephemerides binned by heliocentric distance.  Oort cloud and long-period comets have higher proper motions in the inner Solar System than short-period comets, due to their higher eccentricities and the greater potential for high inclinations, including retrograde orbits.  These faster moving comets are, therefore, more likely to exceed a proper motion of 108\arcsec~hr$^{-1}$.  Less than 50\% of the opportunities to observe Oort cloud or long-period comets inside of 1.7~AU are possible, given the current anticipated tracking limit.  Inside of 1.3 AU from the Sun, there is a $\lesssim$25\% probability that any particular comet could be observed.  This limitation can significantly affect science programs requiring the maximum possible spatial resolution in studies of nearby comets.  Any observations dedicated to the near-nucleus environment will have a limited set of objects to consider, if any.  If we artificially increase the non-sidereal tracking rate by a factor of 2 to 216\arcsec~hr$^{-1}$, the situation is substantially improved: the 50\% limit for Oort cloud/long-period comets is reduced to $r_h\approx1.3$~AU and most comets can be observed down to 1.1~AU from the Sun.

In \reftab{comets} we present a select list of targets observable by \jwst, based on: (1) comets with perihelion dates during the first 5 years of the observatory's mission, (2) Centaurs with known or suspected cometary activity, (3) prior and potential spacecraft targets, and (4) known main-belt comets and activated asteroids.  Our choice of Jupiter-family comets is biased towards previous or potential spacecraft targets, and targets with favorable observing circumstances (small comet-observer distance, or otherwise easily observable at perihelion).  This list of comets, as well as undiscovered comets (especially Oort cloud and long-period comets), comets near aphelion, and unanticipated outbursts and fragmentation events will provide many targets of opportunity for \jwst{} observers.

\section{SUMMARY}

We considered simple tools for estimating comet dust and gas brightness based on known comet properties.  We then explored four science themes that \jwst{} will be especially suited to address: (1) For a moderately bright comet, we can expect studies of the main drivers of cometary activity, \hho, \coo, and CO, out to a heliocentric distance of at least 4~AU; \coo{} can be observed even farther, out to at least 7~AU.  (2) We assessed the observatory's potential to detect ice and organics in the comet population.  A survey of surface ice in Centaurs appears feasible for objects as small as 12~km out to 19~AU from the Sun.  Spectroscopy of Jupiter-family comets, targeted near aphelion during periods of inactivity, can be used to determine ice deposition rates (if occurring), and further understand the uniqueness of the recent discovery of organics on the surface of comet 67P/Churyumov-Gerasimenko.  The properties of water ice grains can be measured, and spatially resolved throughout the coma of a moderately bright comet at 4~AU.  Such a study would provide observational constraints on the lifetime of coma water ice, which can in turn be compared to their physical properties derived from light scattering.  (3) Through the fundamental emission band of water at 2.7~\micron{}, \jwst{} will be able to provide direct tests of the cometary nature of main-belt comets.  (4) Spatially resolved dust studies in the mid-IR can be used to examine the physical properties of the grains, and to test the heterogeneity of nuclei through their coma dust properties.  Finally, there will be many opportunities to observe comets during the first five-years of operations, but the proper motion limit significantly affects target availability within 1.7~AU of the Sun.  Improved tracking performance is highly desired to enhance the ability of the community to further studies of the near-nucleus environment of comets at the best-spatial resolutions possible.

The authors thank an anonymous referee for their insightful critique that improved the manuscript.  MSPK acknowledges support for this work
from NASA (USA) grant NNX13AH67G, and CEW acknowledges partial support from grant NNX13AJ11G.  This work is supported at The Aerospace Corporation by the Independent Research and Development program.

This research made use of Astropy, a community-developed core Python package for Astronomy \citep{astropy13}.

%\bibliography{apj-jour,references,extra}
%\bibliographystyle{apj}

\begin{deluxetable}{lccccccl}
  \tabletypesize{\footnotesize}
  \tablecaption{Observed \afrho{} and \efrho{} values, and their
    ratio. \label{tab:ef2af}}
  \tablewidth{0pt}
  \tablehead{
    \colhead{Comet}
    & \colhead{\afrho}
    & \colhead{$\lambda_A$}
    & \colhead{\efrho}
    & \colhead{$\lambda_\epsilon$}
    & \colhead{$T/T_{BB}$}
    & \colhead{$\epsilon f_{em} / A f_{sca}$}
    & \colhead{Reference\tablenotemark{a}} \\

    & \colhead{(cm)}
    & \colhead{(\micron)}
    & \colhead{(cm)}
    & \colhead{(\micron)}
    & 
    &
  }

  \startdata
  1P/Halley   & 1205 & 1.25 & 3450 & 10.1 & 1.13 & 2.9 & 1985-08-25.6, T86, T88 \\
  1P/Halley   & 5710 & 1.25 & 15400 & 10.1 & 1.13 & 2.7 & 1985-12-13.3, T86, T88 \\
  73P-C/S-W 3 & 243 & 0.7 & 1000 & 13 & 1.07 & $<4.1$\tablenotemark{b} & S06, S11 \\
  73P-C/S-W 3\tablenotemark{c} & 1400 & 1.0 & 4150 & 4.0 & 1.12 & 3.0 & S11 and this work \\
  103P/Hartley 2 & 233 & 0.445 & 970 & 13 & 1.08 & 4.2 & S10, M11 \\
  C/1986 P1 (Wilson)\tablenotemark{d} & 4230 & 1.25 & 9660 & 10.1 & 1.11 & 2.3 & 1987-06-1.2, H89 \\
  C/2012 K1 (Pan STARRS) & 5731 & 0.64 & 14900 & 19.7 & 1.02 & 2.6 & W15 \\
  \enddata

  \tablenotetext{a}{H89 = \citet{hanner89}, M11 = \citet{meech11-epoxi} S06 =
    \citet{sitko06-iauc8717}, S10 = \citet{sitko10-iauc9181}, S11 =
    \citet{sitko11}, S13 = \citet{sitko13-lpsc}, T86 = \citet{tokunaga86-halley-pre}, T88 = \citet{tokunaga88-halley-post}, W15 =
    \citet{woodward15}.}

  \tablenotetext{b}{The observations of 73P-C with the BASS (Broadband
    Array Spectrograph System) indicate the comet's IR
    brightness was increasing on hour timescales, and the \afrho{} and
    \efrho{} values given are based on observations separated by about
    40 min.  Therefore, $\epsilon f / A f$ is potentially an
    upper-limit.}

  \tablenotetext{c}{Values in this row are based on the fit to the
    near-IR spectrum presented in \reffig{sw3}.  The continuum
    temperature scale factor was fixed at $T/T_{BB}=1.12$.}

  \tablenotetext{d}{Old provisional designation: 1986l.}
  
\end{deluxetable}

\begin{deluxetable}{lcccccc}
  \tablecaption{NIRSpec case study: adopted volatile parameters. \label{tab:gas-model}}
  \tablewidth{0pt}
  \tablehead{
    \colhead{Species}
    & \colhead{Band}
    & \colhead{$\lambda_c$}
    & \colhead{Area}
    & \colhead{$\log{\left<Z(\mathrm{1.5 AU})\right>}$}
    & \colhead{$\log{\left<Z(\mathrm{10 AU})\right>}$} \\

    & 
    & \colhead{(\micron)}
    & \colhead{(km$^2$)}
    & \colhead{(molec.\,s$^{-1}$\,cm$^{-2}$)}
    & \colhead{(molec.\,s$^{-1}$\,cm$^{-2}$)} \\
  }

  \startdata

  \hho & $\nu_3$    & 2.7 & 6.0   & 17.15 &  8.30 \\
  \coo & $\nu_2$    & 4.3 & 0.26  & 17.51 & 14.84 \\
  CO   & $\nu(1-0)$ & 4.7 & 0.080 & 18.03 & 16.36 \\

  \enddata

  \tablecomments{$\lambda_C$ is the approximate band center, \citep{crovisier83}, Area is effective active area, and $\left<Z\right>$ is the mean sublimation rate per unit area, given at two heliocentric distances.}

\end{deluxetable}

\begin{deluxetable}{rrrrr
    rrr
    @{\extracolsep{4pt}}r@{\extracolsep{0pt}}rr
    @{\extracolsep{4pt}}r@{\extracolsep{0pt}}rr
  }
  \tabletypesize{\footnotesize}
  \rotate

  \tablecaption{NIRSpec case study: Production rates, total band fluxes, and integration times.} \label{tab:gas-flux}

  \tablewidth{0pt}
  \tablehead{

    & 
    & 
    & 
    & 
    & \multicolumn{3}{c}{\hho}

    & \multicolumn{3}{c}{\coo}

    & \multicolumn{3}{c}{CO}

    \\\cline{6-8}\cline{9-11}\cline{12-14}

    \colhead{$r_h$}
    & \colhead{$\Delta$}
    & \colhead{$E_\sun$}
    & \colhead{$\rho$}
    & \colhead{\afrho}
    & \colhead{$\log{Q}$}
    & \colhead{$\log{F}$}
    & \colhead{$t_{10}$}
    & \colhead{$\log{Q}$}
    & \colhead{$\log{F}$}
    & \colhead{$t_{10}$}
    & \colhead{$\log{Q}$}
    & \colhead{$\log{F}$}
    & \colhead{$t_{10}$} \\

    \colhead{(AU)}
    & \colhead{(AU)}
    & \colhead{($\degr$)}
    & \colhead{(km)}
    & \colhead{(cm)}
    & \colhead{(s$^{-1}$)}
    & \colhead{(W\,m$^{-2}$)}
    & \colhead{(s)}
    & \colhead{(s$^{-1}$)}
    & \colhead{(W\,m$^{-2}$)}
    & \colhead{(s)}
    & \colhead{(s$^{-1}$)}
    & \colhead{(W\,m$^{-2}$)}
    & \colhead{(s)}
  }

  \startdata
9.90 & 9.20 & 133 & 2668 &  20 & 19.18 & -26.64 & \nodata & 24.29 & -20.74 & \nodata & 25.27 & -20.86 & \nodata \\
8.04 & 8.04 &  87 & 2330 &  31 & 21.32 & -24.31 & \nodata & 24.90 & -19.93 &   12000 & 25.45 & -20.48 & \nodata \\
7.05 & 6.55 & 116 & 1901 &  40 & 22.54 & -22.91 & \nodata & 25.16 & -19.50 &    2110 & 25.57 & -20.19 &   59300 \\
5.00 & 4.90 &  90 & 1421 &  80 & 25.26 & -19.84 &   28200 & 25.69 & -18.62 &     127 & 25.87 & -19.54 &    4100 \\
4.01 & 3.63 & 106 & 1052 & 124 & 26.23 & -18.60 &     306 & 25.95 & -18.08 &    32.7 & 26.06 & -19.08 &     841 \\
1.92 & 1.68 &  87 &  488 & 540 & 27.62 & -16.39 &   0.731 & 26.69 & -16.53 &   0.942 & 26.70 & -17.62 &    52.4 \\
1.47 & 1.08 &  90 &  313 & 925 & 27.94 & -15.71 &   0.148 & 26.93 & -15.92 &   0.288 & 26.94 & -17.02 &    36.8 \\
2.39 & 2.16 &  91 &  627 & 350 & 27.33 & -16.94 &    2.79 & 26.48 & -16.99 &    2.56 & 26.51 & -18.06 &    81.2 \\
3.03 & 2.65 & 102 &  769 & 218 & 26.93 & -17.59 &    14.9 & 26.25 & -17.46 &    7.61 & 26.31 & -18.51 &     186 \\
5.19 & 5.10 &  89 & 1480 &  74 & 25.00 & -20.14 &   98400 & 25.64 & -18.71 &     163 & 25.84 & -19.61 &    5430 \\
6.05 & 5.41 & 126 & 1569 &  55 & 23.86 & -21.41 & \nodata & 25.42 & -19.05 &     449 & 25.70 & -19.88 &   15400 \\
8.09 & 8.03 &  90 & 2329 &  31 & 21.26 & -24.37 & \nodata & 24.88 & -19.95 &   13200 & 25.45 & -20.49 & \nodata \\
9.04 & 8.50 & 120 & 2464 &  24 & 20.12 & -25.61 & \nodata & 24.56 & -20.38 &   84600 & 25.35 & -20.69 & \nodata \\

  \enddata

  \tablecomments{As an approximation, an Earth-based observer viewing comet C/2013 A1 (Siding Spring) was used to generate observer-comet distances, $\Delta$, and solar elongations, $E_\sun$. $\rho$ is the radius of a 0\farcs4 synthetic aperture projected at the distance of the comet, \afrho{} is the parameter of \citet{ahearn84-bowell}, $Q$ is the production rate, $F$ is the total band flux, $t_{10}$ is the integration time to achieve a signal-to-noise ratio of 10.}

\end{deluxetable}

\begin{deluxetable}{ccccc}
  \tabletypesize{\footnotesize}

  \tablecaption{Comet 46P/Wirtanen surface brightness and MIRI MRS IFU sensitivities. \label{tab:miri-est}}

  \tablewidth{0pt}
  \tablehead{
    \colhead{$\lambda$}
    & \colhead{FOV}
    & \colhead{$I_\lambda$} 
    & \colhead{$\sigma$}
    & \colhead{SNR} \\

    \colhead{(\micron)}
    & \colhead{(\arcsec)}
    & \colhead{($10^{-5}$~W~m$^{-2}$~\micron$^{-1}$~sr$^{-1}$)}
    & \colhead{($10^{-5}$~W~m$^{-2}$~\micron$^{-1}$~sr$^{-1}$)}

  }

  \startdata
   6.4 & $0.18\times0.19$ & 1.1 & 0.11 &  10 \\
   9.2 & $0.28\times0.19$ & 2.9 & 0.05 &  60 \\
  14.5 & $0.39\times0.24$ & 3.2 & 0.01 & 320 \\
  22.5 & $0.64\times0.27$ & 1.6 & 0.02 &  80 \\
  \enddata

  \tablecomments{Computed for 2019 March 13 ($r_h=1.56$~AU, $\Delta=0.69$~AU) in a 0\farcs4 radius aperture and 100~s of integration time per module.   $\lambda$ is the central wavelength of the estimate, FOV is the field of view of the instrument's native resolution element, $\sigma$ is the estimated 1-$\sigma$ continuum sensitivity, $I$ is the modeled continuum surface brightness, and SNR is the signal-to-noise ratio per resolution element.}

\end{deluxetable}

\begin{deluxetable}{lcp{3in}}
  \tabletypesize{\footnotesize}
  \tablecaption{Potential target comets during the first 5 years of
    \jwst. \label{tab:comets}}
  \tablewidth{0pt}
  \tablehead{
    \colhead{Comet}
    & \colhead{$T_P$}
    & \colhead{Notes}
  }

  \startdata
  2P/Encke                         & 2020-06-26 & Potential spacecraft target\\
                                   & 2023-10-23 & \\
  6P/d'Arrest                      & 2021-09-18 & Potential spacecraft target\\
  21P/Giacobini-Zinner             & 2018-09-10 & Spacecraft target\\
  29P/Schwassmann-Wachmann 1       & 2019-03-07 & Centaur, frequent strong outbursts\\
  38P/Stephan-Oterma               & 2018-11-10 & Halley-type \\
  46P/Wirtanen                     & 2018-12-12 & Potential spacecraft target; historic close approach to Earth (0.078 AU)\\
  64P/Swift-Gehrels                & 2018-11-03 & Excellent apparition\\
  67P/Churyumov-Gerasimenko        & 2021-11-02 & Spacecraft target\\
  95P/Chiron (2060)                & 2046-08-03 & Centaur with jets/arcs/cometary activity \citep{bus91, elliot95, ruprecht15} \\
  99P/Kowal 1                      & 2022-04-12 & Large $q$ (4.7 AU)\\ 
  103P/Hartley 2                   & 2023-10-12 & Spacecraft target\\
  104P/Kowal 2                     & 2022-01-08 & Excellent apparition\\
  107P/Wilson-Harrington           & 2022-08-25 & Active asteroid with small $q$ (0.97 AU)\\
  117P/Helin-Roman-Alu 1           & 2022-07-08 & Large $q$ (3.0 AU)\\
  126P/IRAS                        & 2023-07-05 & Halley-type orbit \\
  176P/LINEAR 52                   & 2022-11-21 & MBC active by sublimation\\
  133P/Elst-Pizarro                & 2018-09-21 & MBC active by sublimation\\
  289P/Blanpain                    & 2019-12-20 & Nearly extinct comet \citep{jewitt06}; historic close approach to Earth (0.086 AU)\\
  311P/PanSTARRS 23                & 2020-10-07 & MBC active by rotational breakup \citep{moreno14}\\
  P/2010 R2 (La Sagra)             & 2021-05-09 & MBC active by sublimation \citep{hsieh12-lasagra}\\
  P/2013 R3 (Catalina-PANSTARRS 1) & 2018-12-06 & MBC active by rotational breakup \citep{jewitt13}\\
  C/2010 U3 (Boattini)             & 2019-02-26 & Oort cloud comet with very large $q$ (8.5 AU)\\
  C/2014 F3 (Sheppard-Trujillo)    & 2021-05-23 & Unusual orbit: low inclination and long (60-year) period \\
  (596) Scheila                    & 2022-05-26 & MBC active by impact \citep{jewitt11-scheila, bodewits11-scheila}\\
  (3200) Phaethon                  & 2019-07-03 & Active asteroid with small $q$ (0.14 AU) \citep{li13-phaethon} \\
  (10199) Chariklo                 & 2066-06-26 & Centaur with rings \citep{braga-ribas14} \\
  \enddata

  \tablecomments{$T_P$ is the comet's perihelion date, MBC = main
    belt comet, $q$ is perihelion distance.}

\end{deluxetable}
\clearpage

\end{document}